\documentclass[]{appolb}
\usepackage[sort&compress,square,comma]{natbib}
\usepackage{amssymb}
\usepackage{graphicx}
\def\Ener{\cal E}
\def\Press{\rm P}
\def\nb{n_{\rm \scriptscriptstyle b}}
\def\ntr{n_{\rm \scriptscriptstyle 0}}
\def\gamN{\gamma_{\rm \scriptscriptstyle N}}
\def\gamm{\gamma_{\rm \scriptscriptstyle m}}
\def\mN{m_{\rm \scriptscriptstyle N}}
\def\mm{m_{\rm \scriptscriptstyle m}}
\def\K{\rm K}
\def\KN{\rm K_{\rm \scriptscriptstyle N}}
\def\Km{\rm K_{\rm \scriptscriptstyle m}}
\def\rhoN{\rho_{\rm \scriptscriptstyle N}}
\def\rhoS{\rho_{\rm \scriptscriptstyle S}}
\def\rcorem{r_{\rm \scriptscriptstyle m}}
\def\rcoreN{r_{\rm \scriptscriptstyle N}}

\begin{document}
\title{Changes in the global parameters of polytropic stars 
induced by the appearance of the soft core}
\author{M. Bejger
\address{N.\ Copernicus Astronomical Center, \\
	  Bartycka 18, 00-716 Warsaw, Poland \\
          {\tt bejger@camk.edu.pl}}}
\maketitle
\begin{abstract}
The effect of a soft phase core appearance in the center of polytropic star 
is analyzed by means of linear response theory. 
Approximate formulae for the changes of radius, 
moment of inertia and mass-energy of non-rotating 
configuration with arbitrary adiabatic indices 
are presented, followed by an example 
 evaluation of astrophysical observables. 
\end{abstract}

\PACS{04.40.Dg, 97.60.Jd}

\section{Introduction}
\label{sec:intro}
In spite of lack of precise Earth-based experimental
data involving matter at densities higher that the 
nuclear saturation density the astrophysical observations
of compact objects give us an unique chance of understanding
the underlying physics. At present it cannot be excluded
that at certain density phase transition in nuclear matter
will produce a state not observed in laboratories e.g. 
pion or kaon condensation, or de-confined quarks 
(see \citep{weber99} for review). If the phase transition 
happens to be of first order, the consequences from the 
point of view of observations are extremely interesting. 
Namely, in a first order phase transition, the new phase
 arise only by nucleation. This means that the 
meta-stable core formed during the compact star evolution
(e.g. accretion, spin-down) nucleates to a stable new-phase
core. The transition is therefore accompanied by star-quake -- 
the radius change, energy release and possible other violent 
phenomena. 

This work is an extension of the linear response theory
developed by P. Haensel, L. Zdunik and R. Schaeffer 
\citep{zhs_ph87,hzs_ph86} and employed to 
the case of first order phase transition from one pure 
phase (normal, N-phase) to other pure phase (super-dense, 
S-phase). While one demands that the transition should 
conserve {\it locally} electric charge and the number
of particles, it occurs at constant pressure, technically 
by so-called ``Maxwell construction'' \citep{ll93}. In 
the presence of the gravitational field it is 
accompanied by a density jump at the boundary of phases.

Due to the works of N. Glendenning \citep{Glend91,Glend92} 
it became clear that relaxing the condition of local 
electrical neutrality permits for a coexistence of {\it both} 
phases within a range of pressures -- a {\it structural} mixed-phase
transition. The volume fraction
occupied by the higher-density phase increases from 
zero at the lower pressure boundary to one at the upper
pressure boundary. In realistic matter, if the surface tension
and Coulomb contribution to the energy is not to large,
the mixed-phase is preferred over a pure phases state. 
 
Here we derive the formulae suitable
for description of the change of the parameters (radius,
moment of inertia, mass-energy) of a 
compact star under the transition from a meta-stable
core to stable core composed of the ``structural mixed-phase''. 
These changes are proportional to the specific powers of the newly-born
core. We will neglect the 
rotation of the star. For simplicity,
the equation of state (EOS) of matter in both N and the mixed 
phases will be approximated by polytropes. The methods 
used here are similar to those used in a somewhat more complicated 
 case of realistic N-phase EOSs presented in \citep{Bejger2005}. 
 
The article is arranged as follows: 
 Sect.~\ref{sec:linear} provides brief description
of the theoretical background and methods used, 
Sect.~\ref{sec:results} contains results of numerical 
calculations, and Sect.~\ref{sec:conclusions} includes
conclusion and final remarks. 
\section{Linear response theory}
\label{sec:linear}
\begin{figure}[t]\centering
\resizebox{!}{2.75in}{\includegraphics[clip]{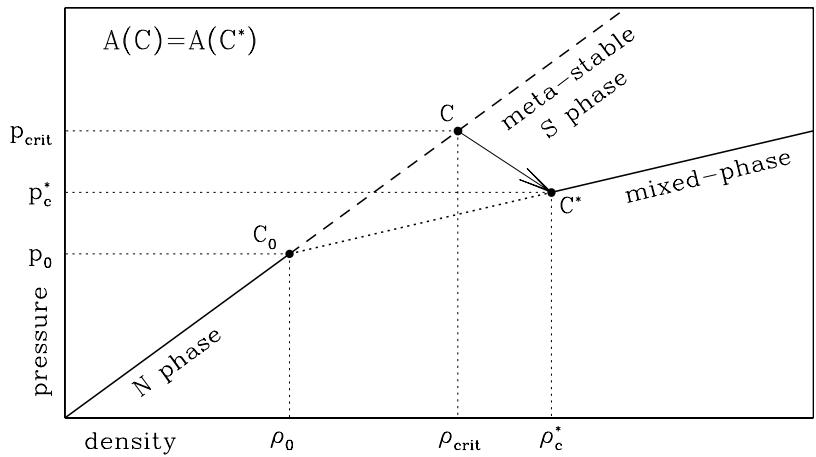}}
\caption{Schematic plot of central pressure  $P_{\rm c}$ versus
 central matter density $\rho_{\rm c}$ for configurations 
based on a pure N phase EOS and an EOS
 with a mixed-phase segment. The solid line denotes stable states,
  the dashed line -- the states which are meta-stable with respect to the
  transition to a mixed-phase state. For a critical central density
$\rho_{\rm crit}$ the S phase nucleates in the super-compressed core
of configuration C, which results in the transition
into a stable configuration C* with a mixed-phase core 
 and a central density $\rho^*_{\rm c}$. 
 Configurations C and C* have the same baryon number $A$.}
\label{fig:pc_rhoc}
\end{figure}

We will describe theoretical background of star linear response 
to the appearance of the soft ``mixed-phase'' core. The
calculation  is based on expressing the change in the density
profile, due to the presence to a small core, as the combination
of two independent solutions of the linearly perturbed equations
of stellar structure \citep{hzs_ph86,zhs_ph87}.
The presence of a denser phase in the core changes the boundary
condition at the phase transition pressure 
${\Press}_0$ (Fig.~\ref{fig:pc_rhoc})
and allows to determine the numerical coefficients in the
expression for the density profile change. The leading term in the
perturbation of the boundary condition at the edge of the new
phase results from the mass excess due to the lower stiffness, and
higher density of the new phase as compared to those of the (normal,
less-dense) N-phase.

We assume that at a central
pressure ${\Press}_{\rm c}={\Press}_{\rm crit}$ the nucleation of the
S-phase in a super-compressed core of radius ${\rcoreN}$, of
configuration ${\rm C}$, initiates  the phase transition and
formation of the ``mixed-phase'' core of radius ${\rcorem}$ in a new
configuration ${\rm C}^*$, as shown on Fig.~\ref{fig:pc_rhoc}. 
Transition to a mixed phase
occurring at ${\rcorem}$ is associated with a substantial drop
in the adiabatic index of matter, defined as $\gamma\equiv
({\nb}/{\Press}){\rm d}{\Press}/{\rm d} {\nb}$, from 
 ${\gamN}$ to ${\gamm}$. In realistic EOSs, mixed phase is 
softer than the pure one, because the increase of mean density 
is reached partly via conversion of a less dense N phase into 
denser S phase, and therefore requires less pressure than 
for a pure phase. This remains true for any fraction of the S phase, 
and leads to a discontinuity of $\gamma$ at the phases boundary, 
$\rho_0$. Realistic examples of such transitions are given by 
\citep{GlendSchaff1998,GlendSchaff1999,Pons2000} and 
\citep[][Table 9.1]{glend.book}. In all those cases,
near the boundary logarithm of pressure depends linearly on the 
logarithm of baryon density; it clearly indicate that the 
polytropic approximation of mixed-phase is valid in the 
small mixed-phase core regime. We will henceforth benefit 
from expansion in powers of the (tiny) core radius ${\rcorem}$.

As far as global properties of stars are concerned,    
the hydrostatic stars' equilibria corresponding to EOSs with 
and without the phase transition must be compared.  
 The models are non-rotating, spherically symmetric
solutions of Einstein's equations, usually called 
Tolman-Oppenheimer-Volkoff (TOV) equations 
 \citep{Tolman39,OppVolk39}. The moment of inertia was 
calculated using the slow-rotation approximation \citep{Hartle1967}. 
 Hydrostatic solutions are labeled by their 
central density $\rho_{\rm c}$.
The configurations based on the two
EOSs are identical up to 
 $\rho_{\rm c}=\rho_0$; 
 the configuration 
 with such central density will be denoted by ${\rm C}_0$ and
will be called a ``reference~configuration''.

As was demonstrated by \citep{Bejger2005}, the leading changes due to
 a phase transition are proportional to the fifth power of 
 ${\rcorem}$ in the case of 
 stellar radius $R$ and the moment of
inertia $I$ and seventh power in the case of 
 the gravitational mass-energy $E=Mc^2$. 
During calculations, we assume constant ${\gamm}$ 
in the ``mixed-phase'' -- in principle, the inclusion of ${\rcorem}$-
dependent ${\gamm}$ contributes to the higher order terms,
but this contribution is negligible in the case of realistic
EOSs (see \citep{Bejger2005} for details). 

 When the central density exceeds
$\rho_0$, the models begin to differ due to the appearance of a 
softer ``mixed-phase'' core in configurations
corresponding to the mixed-phase segment in the EOS.
For two stars composed of equal number of baryons $A$, greater than a
baryon number $A_0$ of the reference configuration ${\rm C}_0$, we 
compare the global parameters -- mass-energy, radius,
 and moment of inertia. Their difference corresponds to the changes in these
 parameters implied by the phase transition in the stellar core.

 The new phase 
 influences the boundary conditions by the
presence of the prefactor $({\gamN}/{\gamm}-1)$,
which is qualitatively the same as $({\rhoS}/{\rhoN}-1)$ 
 in the case of a density jump from pure
N phase to pure S phase \citep{hzs_ph86,zhs_ph87}. Moreover, 
 the results obtained by \citep{Bejger2005} 
 indicate that linear response effects should be 
proportional to the core radius power greater 
 by two than in the case of the transition 
 between pure phases. One would expect that the relative 
 change of the stellar parameter $Q=R,~I,~E$ takes the following form:
\begin{equation}  \label{beta1}
\delta\bar{Q} \equiv \frac{Q^*-Q}{Q_0} \simeq
-\beta_{Q}({\gamN}/{\gamm}-1)({\bar{r}_{\rm\scriptscriptstyle m}})^l,
\end{equation}
where $\bar{r}_{\rm\scriptscriptstyle m}\equiv {\rcorem}/R_0$, 
 the power $l = 5$ corresponds 
 to the radius $R$ and moment of inertia $I$, and $l = 7$ for the mass-energy
$E=Mc^2$. The coefficients $\beta_Q$ are then the 
 functionals of a reference configuration, i.e. 
 $\beta_Q=\beta_Q({\rm C}_0)$. 
The validity of the above statements will be 
 confirmed in the next section  
 by means of numerical calculations.
\section{Results}
\label{sec:results}
In order to obtain quantitative results the
polytropic EOSs for the N and mixed phases will be used. The main
reason why we have chosen the polytropic EOSs are precision
of numerical calculation and simplicity during the exploration
of the parameter space. 
Discussion of the polytropic EOSs and their application to relativistic stellar
structure calculations was presented by \citep{Tooper65}. Details
needed for the calculations are given in the Appendix.
\begin{figure}[t]\centering 
\resizebox{!}{2.75in}{\includegraphics{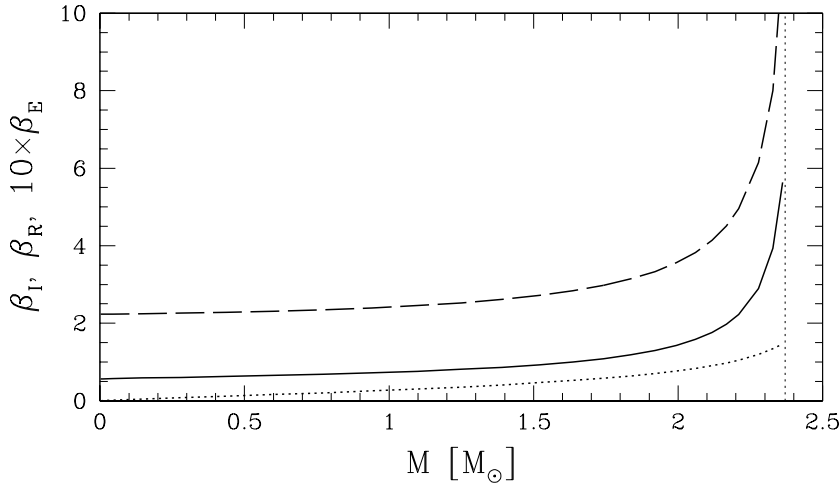}}
\caption{The response coefficients $\beta_I$, $\beta_R$, 
$\beta_E$ (dashed, solid and dotted line, the value 
of $\beta_E$ was multiplied by
a factor of 10) plotted against the stellar mass for a sample
polytropic EOS, ${\gamN}=2.5$, ${\gamm}=1.5$,
${\KN} =0.025$ (see the Appendix for details).
The vertical dotted line denotes the maximal
mass of the star, $2.37~M_{\odot}$.} \label{fig:betavsmass}
\end{figure}
The Sect.~\ref{sec:linear} described 
 the form of the expressions for the changes of
stellar parameters obtained in the linear approximation.
According to Eq.~(\ref{beta1}), the coefficients split into two
factors, first one depending on the mixed phase via 
 $\gamma_{\rm \scriptscriptstyle S}$, and the
second one $\beta_Q$ being a functional of the reference configuration 
${\rm C}_0$. As it is a complicated task to write an analytical form of the
functional $\beta_Q({\rm C}_0)$ we will restrict
ourselves to the derivation of approximate, but accurate fitting formulae,
based on the results of precise numerical calculations.

The $\beta_I,\beta_R$ and $\beta_E$ coefficients obtained in the limit 
of vanishing mixed-phase core ${\rcorem}\longrightarrow 0$ as a function of 
the mass of the sample reference configuration $M({\rm C}_0)$ 
are plotted in Fig.~\ref{fig:betavsmass}. 
Whole range of masses of the reference configuration
is presented on account of our ignorance about the density at which the
phase transition takes place. Results were obtained for specific choices of
${\gamN}$ and ${\gamm}$. The values of ${\gamN}$ and 
 ${\KN}$ were chosen in such a way, that the EOS of the
N-phase produced massive ($M \gtrsim {\rm M}_\odot$) neutron-star
models similar to those obtained for realistic stiff EOS of dense
matter. In particular, this EOS yields $M_{\rm max}=2.37~{\rm
M}_\odot$ and $R_{M_{\rm max}}=12.52$ km. Let us
stress, however, that while this EOS is a reasonable
representation of the EOS of matter with $\rho >2\rho_0$, it is
completely unrealistic at sub-nuclear densities and for masses 
much smaller than one solar mass. 

One notices a characteristic behavior of  $\beta_R$ and $\beta_I$.
Despite the fact that we do not know the precise density of
onset of the new phase, this information is not substantial to 
predict stars response. The coefficients are almost constant for a wide 
range of masses for $\beta_I$ and $\beta_R$. 
The behavior of $\beta_E$ is different: for
$M \lesssim 0.8M_{\rm max}$ this coefficient is proportional to 
$M({\rm C}_0)$. 

The ``plateau'' values of the coefficients $\beta_I$ and
$\beta_R$ do not depend on the pressure
coefficient ${\KN}$, but on ${\gamN}$ only. The 
 formulae which describe accurately those response 
 parameters read 
\begin{equation}\label{betaRI}
\beta_{R}({\rm C}_0) \simeq  
\frac{0.015\times{\gamN}^{9.4}}{({\gamN}^{1.13}-1)^{8.2}}~,~~~~~
\beta_{I}({\rm C}_0) \simeq  
\frac{0.12\times{\gamN}^{9.1}}{({\gamN}^{1.22}-1)^{7.5}}~,
\end{equation}
for $\beta_{R}$ and $\beta_I$, respectively.

The case of the mass-energy parameter $\beta_E$ is different, as
it can be seen on Fig.~\ref{fig:betavsmass}. As we mentioned before,
sufficiently far from the maximal mass, it is proportional to
$M({\rm C}_0)$. We can approximate the value of
$\beta_E({\rm C}_0)$ as follows:
\begin{equation}\label{betam}
\beta_{E}({\rm C}_0)\simeq \frac{0.085}
{{\gamN}^{2.56}{\KN}^{0.5({\gamN}-1)}} 
\times\left(\frac{M}{M_{\odot}}\right)~.
\end{equation}
It has to be mentioned that the fitting formulae from 
 Eqs.~(\ref{betaRI}--\ref{betam}) have been checked against the adiabatic
 indices in the range 5/3 to 3.5. 
The fitted expressions are fairly accurate within
 a wide range of masses, to within few per-cent compared to 
the exact numerical calculations. 
The response coefficients underestimate 
the magnitude of the linear response for configurations near the 
maximum allowable mass, $M_{\rm max}$.
  The increase of $\beta_Q$ for $M({\rm C}_0)\longrightarrow M_{\rm
  max}$ is due to a ``softening'' of the reference configuration by the
   effects of General Relativity (similar to the case presented in  
 \citep{zhs_ph87,hzs_ph86}). 
Thus the linear approximation and, what follows, the approximate
   expressions presented above cease to be valid in this region.

It should be also expected, as far as the realistic EOS of 
 the crust is concerned, 
 that the assumption of constant $\gamma$ in the crust 
is an over-simplification of the 
problem. Therefore, the region of small masses 
 (smaller than $0.5~M_{\odot}$)
is affected by this unrealistic type of crust. In the case of realistic 
NSs, the behavior of the $\beta_Q$ coefficients near the star's minimum
mass should, due to the softening of matter, be generally similar to 
those near the maximal mass (such calculations for 
 the case of realistic EOSs 
 are presented in \citep{Bejger2005}). 

The fitting formulae for $\beta_Q$ in the region of
validity of the linear-response approximation allow us to compute 
the change of the interesting stellar parameters for every
pair of the parameters ${\gamN}$ and ${\gamm}$. For
example, the $1.4~M_{\odot}$ ${\rm C}_0$ 
configuration from Fig.~\ref{fig:betavsmass} 
has radius $R_0=15.18~{\rm km}$ and moment of inertia 
$I_0=2.27\times 10^{45}~{\rm g\cdot cm^2}$. The response coefficients
are then equal $\beta_R=0.62$, $\beta_I=2.23$ and $\beta_E=0.03$.
The appearance of ${\rcorem}=1$ km core with ${\gamm}=1.5$ adiabatic  
 index inside the ${\gamN}=2.5$ polytrope star changes the radius 
 by $\sim~1$ cm, a number which can be related to 
the change of radius during macro-glitches
in pulsar timing. A relative change of the moment of inertia 
($\Delta I/I=-\Delta \Omega/\Omega$) implies speed-up 
of the order of $10^{-6}$, again a value common to pulsar 
macro-glitches. The released energy equals $3\times 10^{44}~{\rm erg}$.
If the core radius is $4~{\rm km}$ (still well described by 
the linear response theory) the change in radius is $8~{\rm m}$,
an impressive number even in view of the size of terrestrial earthquakes. 
 The value of $\Delta\Omega/\Omega$ will be 
 equal approximately $10^{-3}$, 
three orders of magnitude larger than in biggest macro-glitches.
 Finally, the creation of a $4~{\rm km}$ ``mixed-phase'' core produces 
$\sim 5\times 10^{48}~{\rm erg}$ of energy. 
\section{Conclusions}
\label{sec:conclusions}
The appearance of soft core was studied by means of the linear
response theory to the star parameters. Simple model of a polytropic
star with the ``mixed-phase'' core provides a set of well-approximated
formulae, usable for estimation of the change of stellar parameters
(radius, moment of inertia, emitted energy) for a given ``stiffness''
i.e. the adiabatic index $\gamma$ of a star, the newly-born core,
and its {\it a priori} unknown radius. 
 The fitted formulae are precise to within a few per-cent. 
Resulting parameter changes are of the order of observed 
astrophysical phenomena, which gives hope for future observations
of phase transitions inside real neutron stars. 
Inclusion of a varying adiabatic index certainly modifies 
 the values of the response coefficients, presented in 
 \citep{Bejger2005}, but the comparison with the results shown  
 here proves that the difference is not dramatic. 

\section*{Acknowledgments} I warmly thank prof. P. Haensel and 
dr L. Zdunik for reading 
a preliminary version of the manuscript and for valuable comments. 
This work was partially supported 
by the KBN grant no. 1P03D-008-27 and by the 
Astrophysics Poland-France (Astro-PF) program.
\section*{Appendix -- relativistic polytropes}
\label{appendix}
The relativistic polytrope is defined 
 as the power-law dependence between pressure 
${\Press}$ and baryon number density ${\nb}$ (see, e.g. \citep{Tooper65}):
\begin{equation}\label{eeospolyp} 
{\Press}({\nb})={\K} n^{\gamma}_{\scriptscriptstyle b},
\end{equation}
where $\gamma$ is often called adiabatic index, and ${\K}$ 
is the pressure 
coefficient\footnote{The coefficient ${\K}$ is expressed in
$\hat{\rho}c^2/\hat{n}^{\gamma}$ units,
where $\hat{\rho}:=1.66\times10^{14}~{\rm g/cm^3}$,
and $\hat{n}:=0.1~{\rm fm^{-3}}$.}.

 Assume that matter is strongly degenerate, so that $T=0$ 
approximation is valid.
For simplicity, we consider one type of baryons in the
outer N phase, with the rest mass ${\mN}=1.66\times 10^{-24}$ g.
The total mass-energy density ${\Ener}$ of particles of 
rest mass ${\mN}$ in N phase is related to 
their baryon number density ${\nb}$ by the First Law of
Thermodynamics
\begin{equation}
\label{eeospolye}
{\Ener}({\nb}) = \frac{\KN}{{\gamN}-1}n^{\gamN}_{\scriptscriptstyle b}
+ {\mN}c^2 {\nb}~.
\end{equation}

Baryon chemical potential match to the change of the energy
of matter, at constant ${\Press}$ and $T=0$, due to an 
increase of the baryon number by one. This implies
\begin{equation}\label{chem_pot}
\mu({\nb})={d{\Ener}\over d{\nb}} =
 {{\Press}+{\Ener}\over {\nb}} =
 {{\KN}{\gamN} \over {\gamN}-1}
 n^{\gamN-1}_{\scriptscriptstyle b} + {\mN}c^2.
 \end{equation}
At zero pressure (the surface of the star) the chemical
potential $\mu$ is equivalent to particle rest energy.

The connection between two polytropes (the one corresponding 
to the core being softer than the outer part, ${\gamm}<{\gamN}$, 
see Fig.~\ref{fig:pc_rhoc} and discussion) aims for 
approximating the transition to the mixed-phase, 
known from many realistic calculations 
\citep{GlendSchaff1998,GlendSchaff1999,glend.book,Pons2000}. 
Thermodynamic equilibrium implies that the choice 
of parameters ${\Km}$, ${\gamm}$, and the {\it mean mass} ${\mm}$  
should assure the continuity of pressure and baryon chemical potential 
are continuous along the transition point ${\ntr}$, that is
\begin{equation}\label{eos_sew}
{\Km} = {\KN}n^{{\gamN}-{\gamm}}_{\scriptscriptstyle 0},~~~~~
{\mm} = {\mN} - {({\gamN}-{\gamm}){\Press}({\ntr})
\over({\gamN}-1)({\gamm}-1){\ntr}c^2}
\end{equation}
\bibliographystyle{polonica}
\bibliography{polycc}
\end{document}